\documentclass[12pt]{article}

\usepackage{amsmath}
\usepackage{amssymb}
\usepackage{graphicx}
\usepackage{psfrag}
\usepackage{color}
\usepackage{sistyle}

\DeclareTextSymbol{\degre}{T1}{6}
\DeclareTextSymbol{\degre}{OT1}{23}

\begin{document}

\title{\bf Single beam interferometric angle measurement}

\author{P. Paolino, L. Bellon\\
\'Ecole Normale Sup\'erieure de Lyon, Laboratoire de Physique\\
C.N.R.S. UMR5672 \\
46, All\'ee d'Italie, 69364 Lyon Cedex 07, France\\}

\maketitle

\noindent Final draft of \textit{Optics Communication}, \textbf{280} (1), pp. 1--9 (2007). \\
\copyright \ 2007 Elsevier B.V. All rights reserved. \\
The original publication is available online on the editor's website:

\noindent http://dx.doi.org/10.1016/j.optcom.2007.07.060

\begin{abstract}

We present an application of a quadrature phase interferometer to the measurement of the angular position of a parallel laser beam with interferometric precision. In our experimental realization we reach a resolution of $\SI{6.8e-10}{rad}$ ($\SI{1.4e-4}{\arcsec}$) for $\SI{1}{kHz}$ bandwidth in a $\SI{2e-2}{rad}$ (\ang{1}) range. This alternative to the optical lever technique features absolute calibration, independence of the sensitivity on the thermal drifts, and wide range of measurement at full accuracy.  
\end{abstract}


\newpage

\section{Introduction}

Angle measurement is important in a number of applications, ranging from machine tool operation to calibration of optical prism through astronomic observations. Recently, it's been intensively used in atomic force microscopy detection \cite{Meyer:1988}, or in the quickly growing field of cantilever-based sensing \cite{Lavrik:2004}. Our specific concern is to measure the angular position of a torsion pendulum with the best accuracy achievable: we need a high resolution to resolve its thermal fluctuations in order to test recent fluctuation theorems for out of equilibrium systems \cite{Douarche:2005,Douarche:2006}. This basic metrological task can be performed in many ways with optical methods \cite{Malacara:1975}, using for instance an autocollimator \cite{Handbook-of-Optics}, an interferometric setup (see for example \cite{Malacara:1970}) or an optical lever scheme with an electronic target (such as a segmented photodiode) \cite{Jones:1961,Gustafsson:1994}. A major challenge of these techniques is to allow both a wide range of measurement and a high precision simultaneously. We will focus here on interferometric setups, which are usually restricted to small angle measurements but feature very good accuracy. After a generic introduction to the sensitivity of these techniques, we will present how their range can be greatly expanded without losing in precision using a quadrature phase approach. This novel technique to perform calibrated measurements of the angular position of a parallel laser beam is based on a quadrature phase interferometer \cite{Bellon:2002}, except for the laser beam configuration: in the current setup, a single beam directly enters the calcite prism and the common part of two resulting beam is directly analyzed.

The paper is organized as follows: in section \ref{section:principle} we explain the principle of the technique under a wider approach of interferometric angle measurements, while in section \ref{section:setup} we describe the actual experimental realization and we report the results of the calibration measurements, emphasizing how this technique allows constant recording of tiny rotations independently from thermal drifts. In section \ref{section:noise} we discuss the noise limit of the measurement. In section \ref{section:optical lever} we compare our method to the widespread optical lever technique (based on a detection with a 2 quadrant photodiode), before concluding in section \ref{section:conclusion}.   

\section{Interferometric angle measurement: principle} \label{section:principle}

Let us first discuss the general background of the measurement we want to perform here: we consider a single laser beam with origin O in the $\mathbf{e}_{x},\mathbf{e}_{y}$ plane, and we would like to measure through interferometry its angular direction $\theta$ (the rotation is thus defined around $\mathbf{e}_{z}$). Typically, this corresponds to the situation where one wants to measure the rotation of an object by attaching a mirror to it and illuminating this mirror with a laser beam, O being both on the rotation axis and at the center of the mirror.

\begin{figure}
\begin{center}
	\psfrag{O}[Bl][Bl]{$O$}
	\psfrag{A}[Bl][Bl]{$A$}
	\psfrag{B}[Bl][Bl]{$B$}
	\psfrag{d}[Bl][Bl]{$d$}
	\psfrag{t}[Bl][Bl]{$\theta$}
	\psfrag{Optics}[Bc][Bc]{Optics}
	\psfrag{Sensor}[Bc][Bc]{Sensor}
	\psfrag{Measurement area}[Bc][Bc]{Measurement area}
	\psfrag{Analysis area}[Bc][Bc]{Analysis area}
	\psfrag{k}[Bl][Bl]{$\mathbf{k}$}
	\psfrag{DL}[Br][Br]{$\Delta L (\theta)$}
	\psfrag{x}[Bl][Bl]{$\mathbf{e}_{x}$}
	\psfrag{y}[Bl][Bl]{$\mathbf{e}_{y}$}
	\psfrag{z}[Br][Br]{$\mathbf{e}_{z}$}
	\psfrag{er}[Bl][Bl]{$\mathbf{e}_{/\!/}$}
	\psfrag{et}[Bl][Bl]{$\mathbf{e}_{\bot}$}
\includegraphics{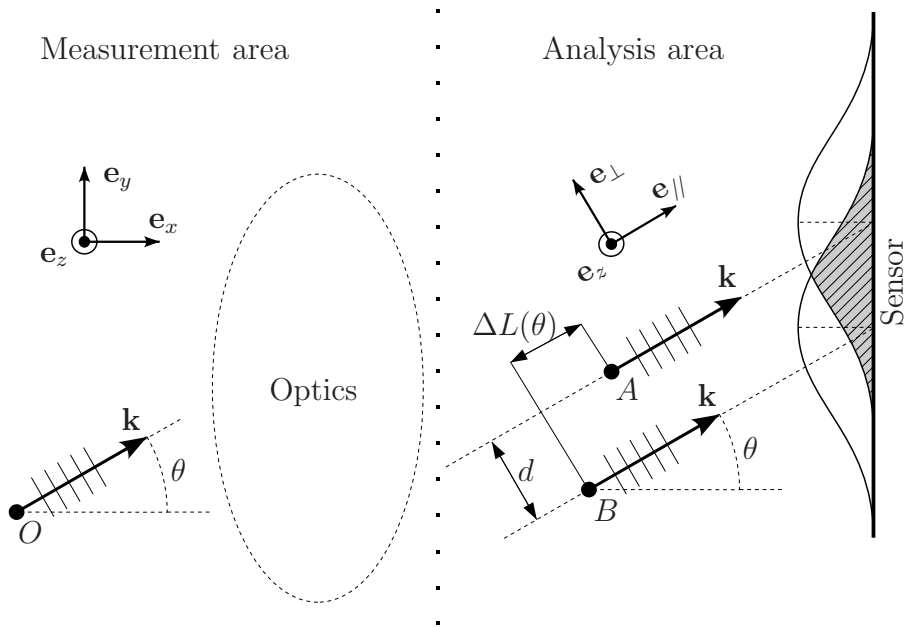}
\end{center}
\caption{Principle of the single beam interferometric angle measurement: the incident light wave is split into two parallel beams in the analyzing region, where they interfere over the sensor. The optical path difference between the 2 beams $\Delta L$ is a function of the angular position of the beam $\theta$.}
 \label{fig:principle}
\end{figure}

In the analyzing area, there should be at least 2 beams in order to have interference. To achieve the best contrast possible, these two beams should have a constant phase difference over the sensor area. This implies in general (given that the sensor is flat and its size is much greater than the wavelength $\lambda$) that the 2 analyzing beams are parallel plane waves over the sensor, and thus in the free space just before it. Their common wave vector is denoted by $\mathbf{k}$ in this area (see Fig.\ref{fig:principle}).

The only assumption underlying the following calculation is that the refractive medium is constant in time (Fermat's principle hypothesis). To simplify the framework we add the assumption that the angular magnification is one, so that the wave vector  $\mathbf{k}$ and its polar angle $\theta$ are shared by the incident light wave and the 2 interfering beams.

Let us now consider the optical path of a ray from the origin O of the incident beam to point A of the first analyzing beam, and an equivalent path from point O to B in the second analyzing beam, such that the optical lengths are equal: $\mathrm{OA=OB}$. That means that as long as A an B are chosen in the free space region before the sensor, the phase difference between the two beams is
\begin{equation}
\varphi (\theta)=\mathbf{k}.\mathbf{AB}=\frac{2 \pi}{\lambda} \Delta L (\theta)
\end{equation}
where $\Delta L (\theta)$ is the optical path difference between the two beams reaching the sensor. We will now demonstrate that this phase difference is a linear function of the angular position $\theta$ of  the incident laser beam.

If we make an infinitesimal change $d\theta$, A will change to $\mathrm{A}'=\mathrm{A}+\mathbf{dA}$ and B to $\mathrm{B}'=\mathrm{B}+\mathbf{dB}$, where we still impose that $\mathrm{OA'=OA=OB'=OB}$. The corresponding phase variation between the two beams is thus
$$d\varphi=\mathbf{dk}.\mathbf{AB}+\mathbf{k}.\mathbf{dAB}$$
\begin{equation}
d\varphi=\mathbf{dk}.\mathbf{AB}+\mathbf{k}.\mathbf{dB}-\mathbf{k}.\mathbf{dA}
\end{equation}
According to Fermat's principle, the optical path is extremal, which implies for free space propagation that the direction of propagation of light is perpendicular to the iso optical length surfaces. It translates here into $\mathbf{k}.\mathbf{dB}=\mathbf{k}.\mathbf{dA}=0$, so that
\begin{equation}
d\varphi=\mathbf{dk}.\mathbf{AB}
\end{equation}
Since the modulus of the wave vector $\mathbf{k}=(2\pi/\lambda) \mathbf{e_{/\!/}}$ is constant, $\mathbf{dk} = (2\pi/\lambda) d\theta \mathbf{e_{\bot}}$ (where $\mathbf{e_{/\!/}}$, $\mathbf{e_{\bot}}$ are polar unity vectors parallel and perpendicular to the propagation). We will eventually only sense rotation of the incident beam with a sensitivity $s$:
\begin{equation} \label{eq:sens}
s=\frac{d\varphi}{d\theta}=\frac{2\pi}{\lambda} d
\end{equation}
where $d=\mathbf{e_{\bot}} \cdot \mathbf{AB}$ is the separation of the beams perpendicular to the propagation (see Fig.\ref{fig:principle}).

According to eq. \ref{eq:sens}, it sounds like there is no limit to the sensitivity, as it is increasing linearly with the distance between the two light rays. However, one should keep in mind that the beams have a finite lateral extension, and to record interference we should overlap them. In gaussian beam approximation, one can show that the optimum sensitivity is achieved when the separation $d$ is equal to the $1/e^{2}$ radius of the beams (see section \ref{section:setup}, eq. \ref{eq:Cmax}).

\section{Experimental setup} \label{section:setup}

\begin{figure}
	\begin{center}
      	   
		\psfrag{D}[Bc][Bc]{\scriptsize D}
		\psfrag{d}[Bc][Bc]{\scriptsize $d$}
		\psfrag{O}[Bc][Bc]{\scriptsize O}
		\psfrag{t}[Bc][Bc]{\scriptsize $\theta$}
		\psfrag{BD0}[Bc][Bc]{\scriptsize BD$_{0}$}
		\psfrag{E0eipy}[Bl][Bl]{\scriptsize $E_0e^{i\varphi}\mathbf{e_y}$}
		\psfrag{E0z}[Bl][Bl]{\scriptsize $E_0\mathbf{e_z}$}
		\psfrag{E0y+E0z}[Bl][Bl]{\scriptsize $E_0\mathbf{e_y}+E_0\mathbf{e_z}$}
		\psfrag{E0eipy+E0z}[Bl][Bl]{\scriptsize $E_0'e^{i\varphi}\mathbf{e_y}+E_0'\mathbf{e_z}$}
		\psfrag{x}[Bl][Bl]{\scriptsize $\mathbf{e_x}$}	
		\psfrag{y}[Bl][Bl]{\scriptsize $\mathbf{e_y}$}
 		\psfrag{z}[Br][Br]{\scriptsize $\mathbf{e_z}$}
		\normalsize
	\includegraphics{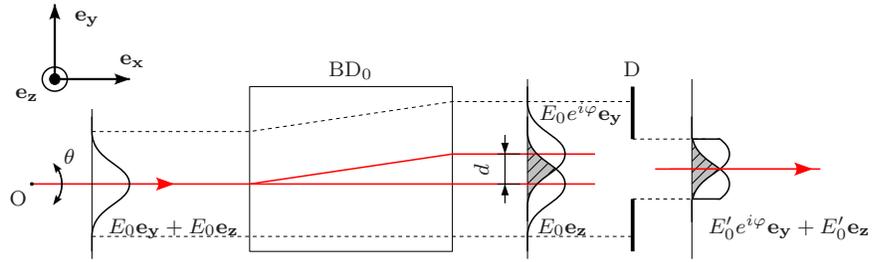}
	\end{center}
\caption{Experimental setup: measurement area. A collimated $\SI{6.6}{mm}$ He-Ne laser beam can be rotated around $\mathbf{e}_{z}$ by means of a piezo driven kinematic mount. After passing through a parallel beam displacer ($\SI{40}{mm}$ calcite prism, BD$_{0}$), the 2 resulting crossed polarized rays present a phase shift $\varphi$ dependent on the angle of incidence $\theta$. We limit the 2 beams to their overlapping part using a diaphragm (D), and analyze the emerging light ray into the analysis area.}
\label{fig:setup-1}
\end{figure}

\begin{figure}
	\begin{center}
      	   
		\psfrag{A1}[Bl][Bl]{\scriptsize $A_1\propto \left|(1 + e^{i\varphi})(\mathbf{e_y}+\mathbf{e_z})\right|^{2}$}
		\psfrag{B1}[Bl][Bl]{\scriptsize $B_1\propto \left|(1 - e^{i\varphi})(\mathbf{e_y}-\mathbf{e_z})\right|^{2}$}
		\psfrag{A2}[Br][Br]{\scriptsize $A_2$}
		\psfrag{B2}[Bl][Bl]{\scriptsize $B_2$}
		\psfrag{PhD1}[Bc][Bc]{\scriptsize PD$_{1}$}
		\psfrag{PhD2}[cr][cr]{\scriptsize PD$_{2}$}
		\psfrag{BD1}[Bc][Bc]{\scriptsize BD$_{1}$}
		\psfrag{BD2}[cr][cr]{\scriptsize BD$_{2}$}
		\psfrag{f}[Bc][Bc]{\scriptsize $f$}
		\psfrag{L1}[Bc][Bc]{\scriptsize L$_{1}$}
		\psfrag{L2}[cr][cr]{\scriptsize L$_{2}$}
		\psfrag{d'}[Bc][Bc]{\scriptsize $d'$}
		\psfrag{E'0xeip+E'0z}[Bc][Bc]{\scriptsize $E'_{0}e^{i\varphi}\mathbf{e_y}+E'_0\mathbf{e_z}$}
		\psfrag{E'0yeip-p/2+E'0z}[Br][Br]{\scriptsize $E'_0e^{i(\varphi-\pi/2)}\mathbf{e_x}+E'_0\mathbf{e_z}$}
		\psfrag{l/4}[cr][cr]{\scriptsize $\lambda/4$}
		\psfrag{x}[Bl][Bl]{\scriptsize $\mathbf{e_x}$}	
		\psfrag{y}[Bl][Bl]{\scriptsize $\mathbf{e_y}$}
 		\psfrag{z}[Br][Br]{\scriptsize $\mathbf{e_z}$}
		\normalsize
	\includegraphics{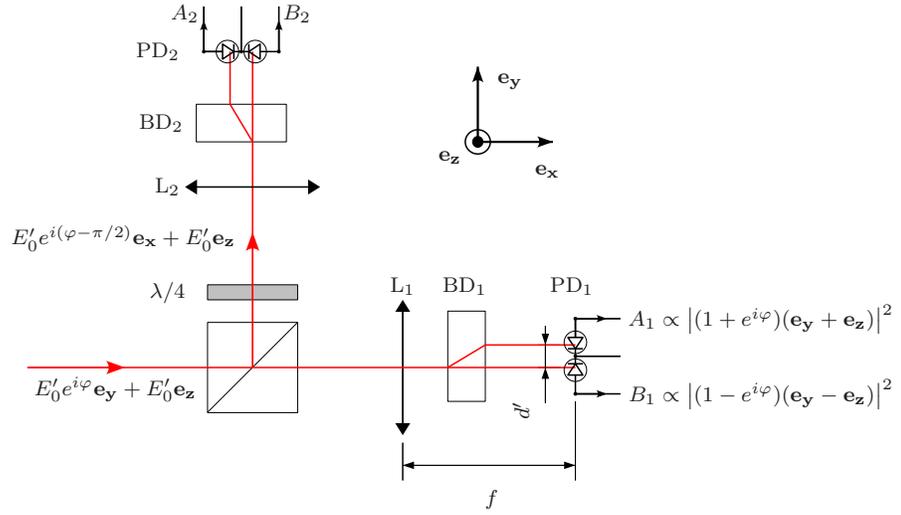}
	\end{center}
\caption{Experimental setup: analysis area. The light coming from the measurement area is split into two arms. In each one, a $\SI{5}{mm}$ calcite prism (beam displacers BD$_{n}$, $n=1,2$) oriented at $\ang{45}$ with respect to the measurement beam displacer (BD$_{0}$) projects the polarizations to have them interfere. The 2 beams emerging BD$_{n}$ are focused (plano convex lens L$_{n}$, $f=25mm$) on the 2 segments of a 2 quadrant photodiode PD$_{n}$ to record their intensities $A_{n}$, $B_{n}$. Those are used to reconstruct $\varphi$ and thus measure $\theta$. In the second analyzing arm ($n=2$), a quarter wave plate ($\lambda/4$) is added in order to subtract $\pi/2$ to the phase shift $\varphi$. }
\label{fig:setup-2}
\end{figure}

We present in this section the experimental setup we have built to demonstrate the workability of an interferometric measurement of the angular position of a single light beam, as schemed on fig. \ref{fig:setup-1} and \ref{fig:setup-2}. The output of a He-Ne laser is sent into a single-mode polarization maintaining fiber, then collimated to a parallel beam of $1/e^2$ diameter $2 R = \SI{6.6}{mm}$. The fiber end and collimator are hold by a kinematic mount with a piezo drive to change the angular direction $\theta$ of the light beam before it enters a parallel beam displacer ($\SI{40}{mm}$ calcite prism, labeled BD$_{0}$ on fig. \ref{fig:setup-1}). We end up with two parallel beams of crossed polarization, separated by $d = \SI{4}{mm}$, thus overlapping a few millimeters. A diaphragm limits the output to this overlapping area. The intensities of the 2 light rays are evenly tuned by adjusting the incident polarization at $\ang{45}$ with respect to the calcite optical axes. No interference can be seen at this stage since the two beams have crossed polarizations, though they present a phase shift $\varphi$ dependent on the angle of incidence $\theta$ of the initial beam on the calcite: Eq. \ref{eq:sens} can directly be used to compute $\varphi$ as a function of $\theta$, since all the hypotheses of section \ref{section:principle} are met (still optics, angular magnification one)\footnote{The same formula can be derived directly analyzing the particular design of the optical setup of fig. \ref{fig:setup-1}, using birefringence laws instead of the formalism of section \ref{section:principle}.}.

We use a quadrature phase technique similar to the one of ref. \cite{Bellon:2002} to analyze these overlapping beams. They are first divided into two equivalents rays with a non polarizing cube beamsplitter. In each arm, the beam is focused ($f=\SI{25}{mm}$ lenses, labeled L$_{1}$ and  L$_{2}$  on fig. \ref{fig:setup-2}) on the detector through a second parallel beam displacer ($\SI{5}{mm}$ calcite prisms, labeled BD$_{1}$ and BD$_{2}$) which optical axis is oriented at $\ang{45}$ with respect to the first calcite BD$_{0}$. We project this way the two initial polarizations and make the two incident beams interfere: the intensities $A$ and $B$ of the 2 beams emerging the last calcite prism are functions of the phase shift $\varphi$ and can be recorded by two photodiodes. Since the two spots are only $\SI{0.5}{mm}$ distant, we actually use a 2 quadrant photodiode. In the second analyzing arm, a quarter wave plate is added in order to subtract $\pi/2$ to the phase shift $\varphi$ between the two cross polarized beams. In the current setup, the use of beam displacers and 2 quadrant photodiodes instead of Wollaston prisms and distinct photodiodes as in ref. \cite{Bellon:2002} make the realization much more compact though as efficient.

Measured intensities $A_{n}$, $B_{n}$ in the two analyzing arms $n$ (with $n=1,2$) are easily computed as:
\begin{eqnarray}
 A_{n} & = & \frac{I_0}{4} (1 + C_{max} \cos(\varphi+\psi_{n})) \nonumber \\ 
 B_{n} & = & \frac{I_0}{4} (1 - C_{max} \cos(\varphi+\psi_{n}))  \label{eq:intensities}
\end{eqnarray}
where $I_0$ is the total intensity corresponding to the incident light beam\footnote{$I_{0}$ is the electrical intensity defined by $I_{0}=SP$, where $P$ is the incident beam power (in $W$) and $S$ is the responsivity of the photodiodes (in $\SI{}{A/W}$). The $1/4$ factor in the equations accounts for the beam-splitting process (2 final beams in both analyzing arms).}, $C_{max}$ is a contrast factor which accounts for lateral extension of the beams ($C_{max}<1$), and $\psi_1=0$ (first arm, without quarter wave plate) or $\psi_2=-\pi/2$ (second arm, with quarter wave plate). Using home made low noise analog conditioning electronic \cite{Electronics}, we can measure for each arm the contrast function of these two signals:
\begin{equation} \label{eq:realcontrast}
 C_n=\frac{A_{n} - B_{n}}{A_{n} + B_{n}}=C_{max} \cos(\varphi+\psi_{n})
\end{equation}
This way, we get rid of fluctuations of laser intensity, and have a direct measurement of the cosine of the total phase shift for each arm, $\varphi+\psi_{n}$.

Let us rewrite eq. \ref{eq:realcontrast} as:
\begin{equation} \label{eq:complexcontrast}
 C=C_1+i \, C_2=C_{max} \left(\cos(\varphi)+i\sin(\varphi)\right)=C_{max} e^{i\varphi}
\end{equation}
Under this formulation, the advantage of using two analyzing arms instead of one is obvious : it allows one to have a complete determination of $\varphi$ (modulo $2\pi$). In the $(C_1,C_2)$ plane, a measurement will lie on the $C_{max}$ radius circle, its polar angle being the phase shift $\varphi$. The sensitivity $s$ of the measurement, defined by eq. 4, appears this way to be independent of the position on the circle of the measurement, and will be constant even with a slow thermal drift. The use of crossed polarizations for the two interfering beams is a key point of this method, since it allows a post processing of the phase difference (with the quarter-wave plate) to produce the quadrature phase signals.  

The beam separation $d$ is in reality function of the angle of incidence $\theta$, but its variation is small in the full $\theta$ range available: the main limitation to the angle of incidence that can be measured is that each beam emerging the last calcites in the analyzing arms should fall on its respective photodiode quadrant (see Fig.\ref{fig:setup-2}) . Given the focal length of the focusing lenses L$_{n}$ ($f = \SI{25}{mm}$) and the separation of the 2 beams ($d' = \SI{0.5}{mm}$), the range accessible in $\theta$ in our setup is $|\theta|<\theta_{max} = d'/2f = \SI{e-2}{rad}$. Note that this range can be greatly extended if useful, choosing a larger separation of the final beams (using a Wollaston prism and 2 distinct photodiodes for example \cite{Bellon:2002,Schonenberger:1989}). Relative variations of $d$ in a $\SI{2e-2}{rad}$ interval for $\varphi$ are within $0.5\%$ for a normal incidence on the calcite, and can be reduced down to $\num{5e-5}$ for an optimal incidence ($\ang{+15}$ angle with the normal of the surface, where positive angles correspond to the direction of the optical axis)

Eventually, all we need to do is acquire the two contrasts and numerically compute
\begin{equation} \label{eq:phi=arctan}
\varphi=\arg(C)=\arctan(C_2/C_1)
\end{equation}
where the $\arctan$ function is extended to the whole $[-\pi,\pi]$ interval according to the signs of $C_{1}$ and $C_{2}$. Note that if $\varphi$ varies in a larger interval, unwraping is necessary to reconstruct the whole signal. Combining this last equation with eq. \ref{eq:sens}, one eventually gets :
\begin{equation} \label{eq:theta=arctan}
\theta=\frac{\lambda}{2 \pi d}\arctan(C_2/C_1)+\theta_{0}
\end{equation}
with $\theta_{0}$ an integration constant.

\begin{figure}
\begin{center}
\includegraphics{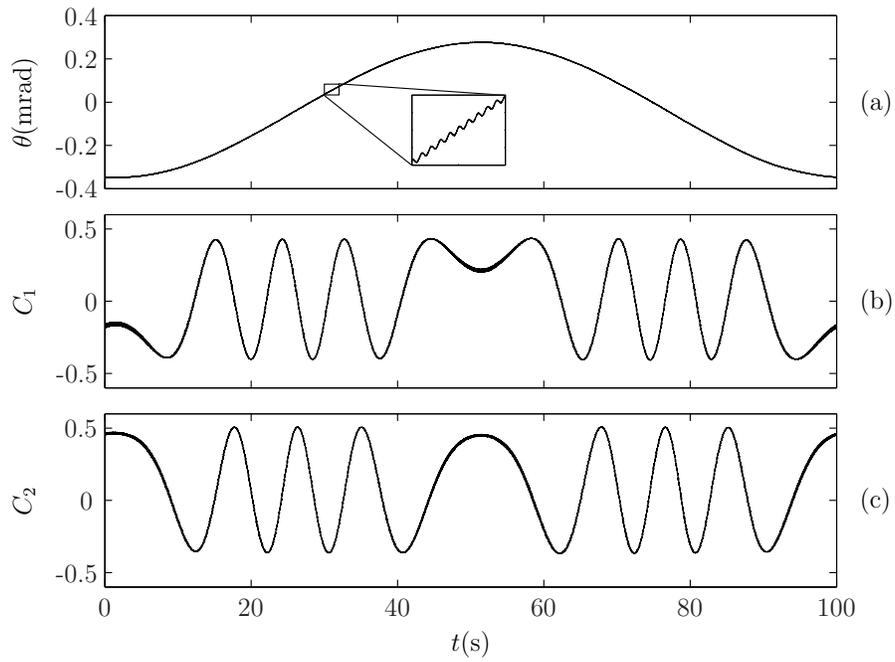}
\end{center}
\caption{(a) Measured angular position $\theta$ of the laser beam. The driving is the sum of two sinusoids : $\SI{0.5}{mrad}$ at $\SI{10}{mHz}$ and $\SI{1}{\micro rad}$ at $\SI{10}{Hz}$. (b) and (c) corresponding contrasts $C_1$ and $C_2$ of the two analyzing arms as a function of time.}
\label{fig:dtheta}
\end{figure}

\begin{figure}
 \begin{center}
\includegraphics{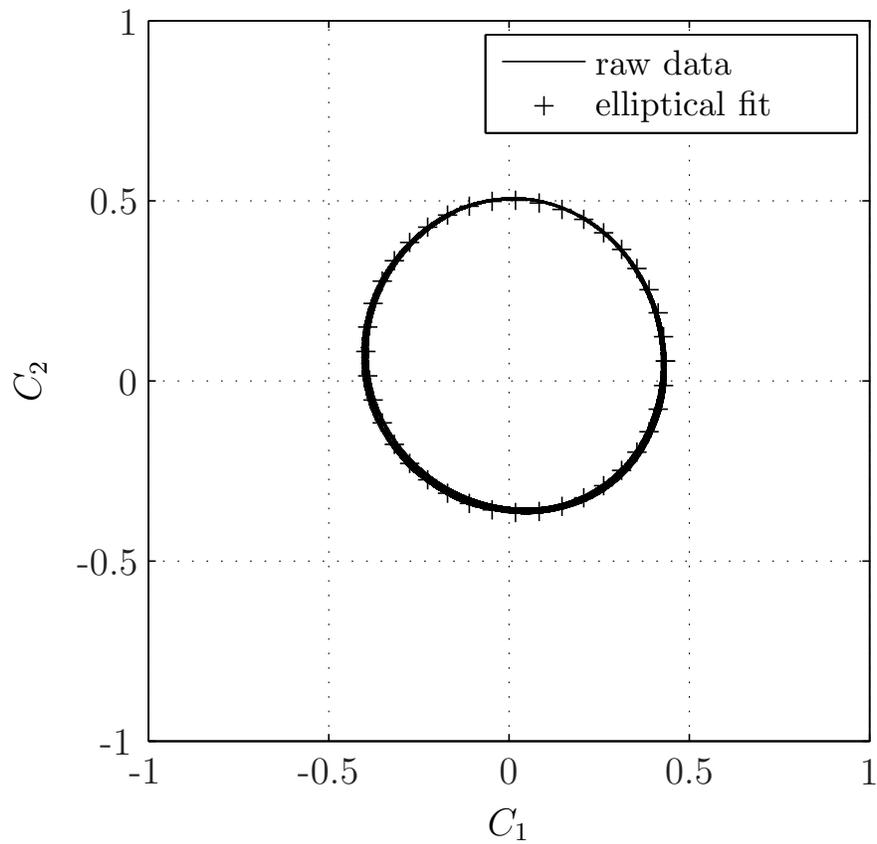}
 \end{center}
\caption{Due to experimental imprecisions, the measurement lays on a tilted ellipse in the $C_1,C_2$ plane. These deviations to the $C_{max}$ radius circle can easily be corrected \cite{Heydemann:1981}. We present the raw data to show that corrections are small anyway.}
 \label{fig:ellipse}
\end{figure}

To demonstrate the operation of this technique, we rotate the beam using a piezoelectric controlled kinematic mount. The driving voltage that we use is the sum of two sinusoids: a fast one of low amplitude (leading to a $\SI{}{\micro rad}$ rotation) and a slow one of high amplitude (simulating a slow drift of the working point of the interferometer over several wavelengths, that is a rotation of about $\SI{1}{mrad}$). In Fig. \ref{fig:dtheta}(a), we plot as a function of time a typical driving of the beam's angular rotation $\theta$. In this specific case the slow and fast sinusoids have a frequency of $\SI{10}{mHz}$ and of $\SI{10}{Hz}$ respectively and the amplitude ratio is about 400. The contrasts $C_1$ and $C_2$ of the two analyzing arms, as expected, are in phase quadrature.  In Fig. \ref{fig:ellipse} we also plot the contrasts $C_1$ and $C_2$ in the $(C_1,C_2)$ plane to show that they lay on a circle.  In fact, the measurement lays on a tilted ellipse, because of the imperfections in the orientation of the beam displacers and quarter wave plate, but this small deviation from the $C_{max}$ radius circle can easily be corrected \cite{Heydemann:1981}. Anyway, as shown by Fig. \ref{fig:ellipse}, the deviations from a circle are small in our setup.

\begin{figure}
 \begin{center}
\includegraphics{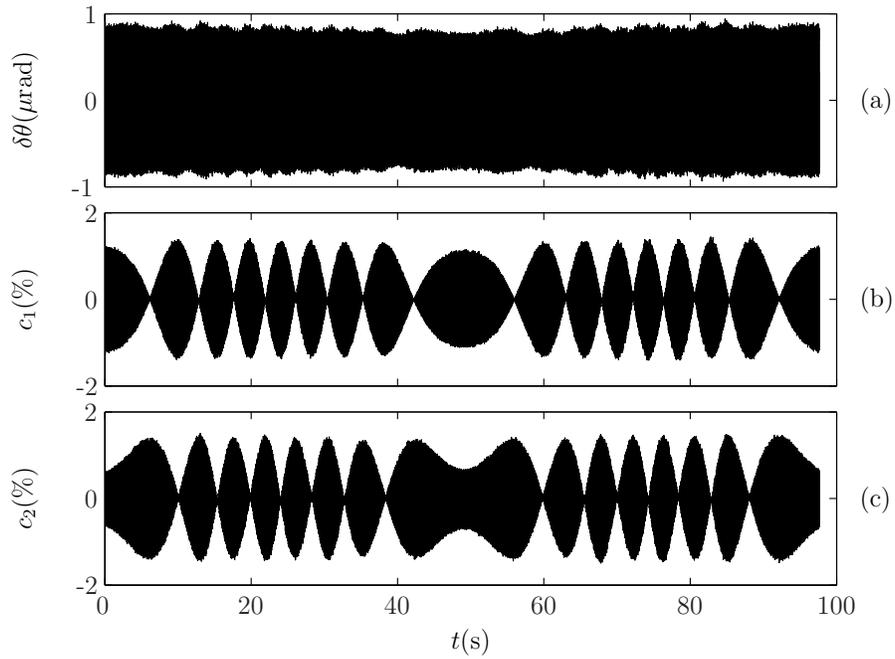}
\end{center}
\caption{Fast evolution of the beam's angular position once the slow variation of Fig.~\ref{fig:dtheta} has been subtracted. (a) Fast angular displacement $\delta\theta$ as function of the time. (b) and (c) Fast contrasts $c_i$ of the two analyzing arms.}
 \label{fig:dtheta_filt}
\end{figure}

\begin{figure}
\begin{center}
\includegraphics{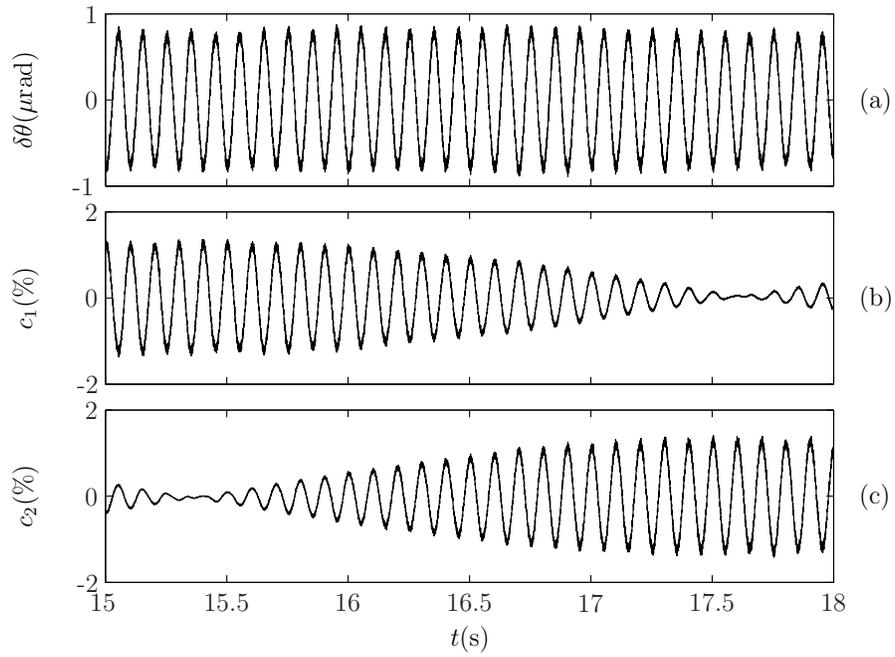}
\end{center} 
\caption{Zoom of Fig.~\ref{fig:dtheta_filt}: fast signals $\delta\theta$ (a), $c_1$ (b), $c_2$ (c) around minimums of  $C_1$ and $C_2$. The reconstructed angular position $\delta\theta$ is independent of the working point.}
 \label{fig:dtheta_filt_zoom}
\end{figure}

Let us now have a closer look at the fast evolution of these signals. In Fig. \ref{fig:dtheta_filt} we plot as a function of time the fast angular displacement $\delta\theta$ and contrasts $c_n$ obtained by a high pass filtration of the signals $\theta$ and $C_{n}$. Comparing Figs. \ref{fig:dtheta_filt}(b) and (c) with Figs. \ref{fig:dtheta}(b) and (c) we see that $c_1(c_2)$ goes to 0 periodically when $C_1(C_2)$ is extremal while the reconstructed angular position has a constant amplitude. Therefore this technique allows constant recording of small rotations as shown in the precedent paragraphs. This is clearly seen in Fig. \ref{fig:dtheta_filt_zoom} where the fast evolution of $\delta\theta$,  $c_1$ and $c_2$ are plotted on a time interval around a minimum of $C_1$ and $C_2$. The cleanness of the curve of Fig. \ref{fig:dtheta_filt_zoom}(a) demonstrates the accuracy of this measurement.

\section{Noise of the measurement} \label{section:noise}

Let us compute the sensitivity $\sigma$ of the complex contrast $C$ as a function of the angle of incidence $\theta$. Using eq. \ref{eq:sens} and \ref{eq:complexcontrast}, we have:
\begin{equation} \label{eq:sens-C/theta}
\sigma = \left |\frac{d C}{d\theta}\right |=C_{max}\frac{2 \pi}{\lambda} d
\end{equation}
$C_{max}$ can be computed analytically in the case of gaussian beams impinging a infinite size sensor. Let us for example consider intensity $A_{1}$:
\begin{equation}  %\nonumber %\label{eq:}
A_{1} \propto \int\!\!\!\!\!\int dy \, dz\, |E_{1}+E_{2}|^2
\end{equation} where
\begin{eqnarray} 
E_{1} & = & E_{0}e^{-\frac{(y-d/2)^2+z^2}{R^2}} \nonumber \\
E_{2} & = & E_{0}e^{-\frac{(y+d/2)^2+z^2}{R^2}}e^{i\varphi} \nonumber %\label{eq:}
\end{eqnarray}
are the electric fields of each beam, and $R$ is their $1/e^2$ radius. It is straightforward to show that
\begin{equation} %\nonumber %\label{eq:}
A_{1} \propto \left(1+ e^{-\frac{d^2}{2R^2}}\cos(\varphi)\right)
\end{equation}
and from eq. \ref{eq:intensities} we directly identify $C_{max}$ as
\begin{equation} \label{eq:Cmax}
C_{max} = e^{-\frac{d^2}{2R^2}}
\end{equation}
The sensitivity $\sigma$ being proportional to $C_{max} d$ (eq. \ref{eq:sens-C/theta}), we can easily show that it is maximum when the separation between the beams is equal to their $1/e^{2}$ radius: $d=R$, where we get $C_{max}=0.61$. In fact, this configuration is not the best one can use to maximize the sensitivity: adding a diaphragm to limit the beams to their common part, we can compute numerically the optimum parameters: $d/R=1.08$ with a diaphragm of diameter $2.25 R$, which lead to $C_{max}=0.65$ and raise the sensitivity $\sigma$ of $15\%$.

The main source of noise in the measurement is the unavoidable shot noise of the photodiodes. Let us denote by $\delta A_{n}$ and $\delta B_{n}$ these shot noise induced fluctuations of $A_{n}$ and $B_{n}$. The power spectrum densities (PSD) of these intensity fluctuations are
\begin{eqnarray}
S_{A_{n}} & = & \left<\delta A_{n}^{2}\right>/\Delta f = 2 e A_{n} \nonumber \\
S_{B_{n}} & = & \left<\delta B_{n}^{2}\right>/\Delta f = 2 e B_{n} \label{eq:shotnoiseAB}
\end{eqnarray}
where $e$ is the elementary charge, $\Delta f$ the bandwidth of the measurement and $\left< . \right>$ stands for time average. They will lead to fluctuations of the contrasts $\delta C_{n}$ which can be written:
\begin{equation}  \label{eq:deltaCn}
\delta C_{n}=\left(\frac{\partial C_{n}}{\partial A_{n}}\right)\delta A_{n} + \left(\frac{\partial C_{n}}{\partial B_{n}}\right)\delta B_{n}
\end{equation}
Since shot noise induced fluctuations $\delta A_{n}$ and $\delta B_{n}$ are uncorrelated, we can compute the PSD of the contrasts using equations \ref{eq:deltaCn}, \ref{eq:realcontrast} and \ref{eq:shotnoiseAB} consecutively:
\begin{eqnarray}  %\nonumber %\label{eq:}
S_{C_{n}} & = & \frac{\left<\delta C_{n}^{2}\right>}{\Delta f} = \left(\frac{\partial C_{n}}{\partial A_{n}}\right)^{2} S_{A_{n}} + \left(\frac{\partial C_{n}}{\partial B_{n}}\right)^{2} S_{B_{n}}  \\
& = & 4 \frac{B_{n}^2}{(A_{n} + B_{n})^4}S_{A_{n}} + 4 \frac{A_{n}^2}{(A_{n} + B_{n})^4}S_{B_{n}} \\
& = & 8 e \frac{A_{n}B_{n}}{(A_{n} + B_{n})^3}
\end{eqnarray}
Using eq. \ref{eq:intensities} and eq. \ref{eq:realcontrast}, we have $A_{n}=I_{0}(1+C_{n})/4$ and $B_{n}=I_{0}(1-C_{n})/4$, hence
\begin{equation} \label{eq:SCn}
S_{C_{n}}= \frac{4 e}{I_{0}}(1-C_{n}^2)
\end{equation}
$\delta C_{1}$ and $\delta C_{2}$ being uncorrelated, we use equations  \ref{eq:phi=arctan}, \ref{eq:SCn} and \ref{eq:realcontrast} consecutively to get the expression of the PSD of the fluctuations of $\varphi$:
\begin{eqnarray}
S_{\varphi} & = & \left(\frac{\partial \varphi}{\partial C_{1}}\right)^2 S_{C_{1}} + \left(\frac{\partial \varphi}{\partial C_{2}}\right)^2 S_{C_{2}} \nonumber \\
& = & \frac{C_{2}^2}{(C_{1}^2+C_{2}^2)^2} S_{C_{1}} + \frac{C_{1}^2}{(C_{1}^2+C_{2}^2)^2} S_{C_{2}} \nonumber \\
& = & \frac{4 e}{I_{0}}\frac{C_{1}^2 (1-C_{2}^2)+C_{2}^2 (1-C_{1}^2)}{(C_{1}^2+C_{2}^2)^2} \nonumber  \\
& = & \frac{4 e}{I_{0}}\left(\frac{1}{C_{max}^{2}}-\frac{1}{2} sin^{2}(2\varphi)\right) %\nonumber %\label{eq:}
\end{eqnarray}
From this last equation and eq. \ref{eq:sens}, we eventually get an upper bound for the power spectrum density of shot noise induced fluctuations in $\theta$:
\begin{equation} \label{eq:maxnoise}
S_{\theta} = \left(\frac{d \theta}{d \varphi}\right)^{2} S_{\varphi} \leqslant \left(\frac{\lambda}{2 \pi d}\right)^2 \frac{4 e}{I_{0}}\frac{1}{C_{max}^{2}} = \frac{4 e}{I_{o} \sigma^2}
\end{equation}

\begin{figure}
 \begin{center}
\includegraphics{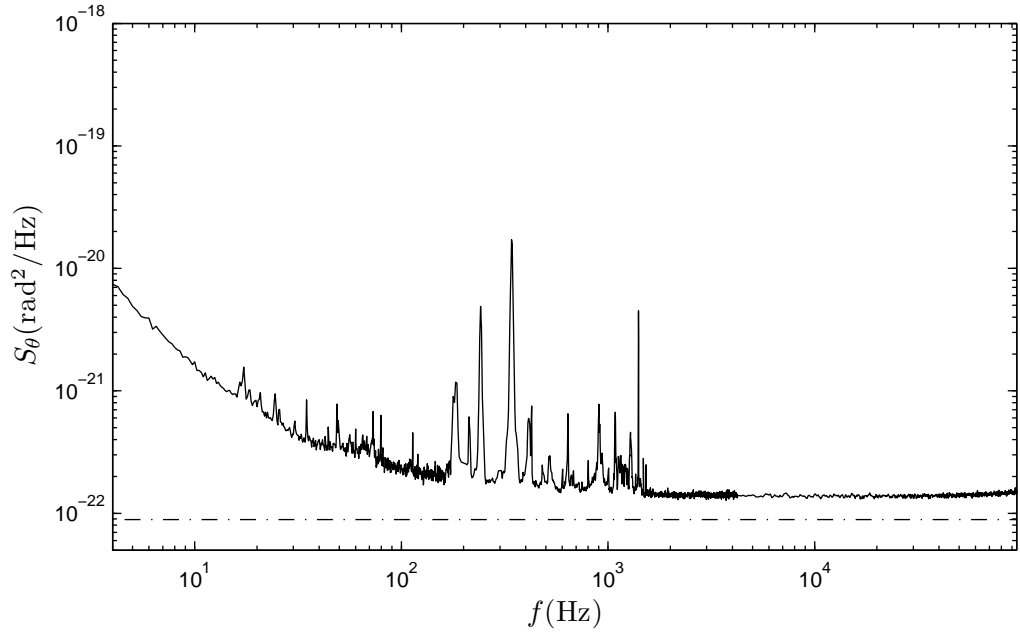}
\end{center}
\caption{Power spectrum density of the angular deflection $\theta$ of a still laser beam (plain line), and maximum shot noise calculated from inequality~\ref{eq:maxnoise} with experimental values of intensity and sensitivity (dash dotted line). The pics in the $\SI{10}{Hz}-\SI{10^3}{Hz}$ region are attributed to mechanical disturbances in the experimental setup and could be addressed by a quieter environment. The integrated noise in the $\SI{0}{Hz}-\SI{1}{kHz}$ range is $\SI{0.68}{nrad}$, which is thus the lower limit of measurable angular displacement for this bandwidth.}
\label{fig:spectre}
\end{figure}

In Fig. \ref{fig:spectre} we plot the power spectrum density $S_{\theta}$ measured with a still laser beam and the shot noise's estimation of our experiment. The pics in the $\SI{10}{Hz}-\SI{10^3}{Hz}$ region are attributed to mechanical disturbances in the experimental setup and could be addressed by a quieter environment, while at low frequency (below $\SI{50}{Hz}$)  the electronics $1/f$ noise is visible. We can see that our setup is close to optimal conditions, with the base line of the noise down to $\SI{1.4e-11}{rad/\sqrt{Hz}}$. Finally we note that  in terms of optical path difference, this value corresponds to a noise of $\SI{5.6e-14}{m/\sqrt{Hz}}$. 

\section{Comparison with optical lever technique} \label{section:optical lever}

In the classic optical lever technique, the single beam illuminates a 2 quadrant photodiode, as sketched in fig. \ref{fig:principle-2Q}. A contrast function $C_{2Q}$ of the intensities of the two quadrants (ratio of the difference and the sum of the signals, similar to the one defined in eq. \ref{eq:realcontrast}) can be used to measure the position of this light beam on the sensor. In appendix \ref{Appendix}, we compute the optimal output for a gaussian beam of $1/e^{2}$ radius $R$ at the center of rotation (eq. \ref{eq:contrast-2Q-Appendix}):
\begin{equation} \label{eq:contrast-2Q}
C_{2Q}=\mathrm{erf}\left(\sqrt{2}\frac{\pi R \sin(\theta)}{\lambda}\right)
\end{equation}
where $\mathrm{erf}$ is the error function. The best sensitivity $\sigma_{2Q}$ is obtained for $\theta \approx 0$ (eq. \ref{eq:sigma-2Q-max}):
\begin{equation} \label{eq:sigma-2Q}
\sigma_{2Q} =  \left(\frac{dC_{2Q}}{d\theta}\right)_{\theta=0} = \sqrt{8 \pi}\frac{R}{\lambda} 
\end{equation}
Given the shape of the $\mathrm{erf}$ function, the range of the measurement is inversely proportional to the sensitivity: $\theta_{max}^{2Q} \sim 1/\sigma_{2Q}$. For a $\SI{7}{mm}$ diameter laser beam at $\SI{633}{nm}$, the admissible range is thus limited to $|\theta|<\SI{2e-5}{rad}$. This range can obviously be extended by degrading the sensitivity (non optimal focusing of the beam).

A computation similar to the one of previous paragraph can be done to analyze the shot noise induced fluctuations in $C_{2Q}$, they result in a power spectral density $S_{C_{2Q}}= 2 e/I_{0}$, which finally leads to
\begin{equation} \label{eq:noise2Q}
S_{\theta}^{2Q}= \left(\frac{\lambda}{R}\right)^2 \frac{1}{8 \pi}\frac{2 e}{I_{0}} = \frac{2 e}{I_{0} \sigma_{2Q}^2} 
\end{equation}
We have supposed up to now a zero width slit for the segmented photodiode. Interestingly, the noise can be reduced by introducing a gap between the quadrants \cite{Gustafsson:1994}. Using optimal separation, the power spectrum density of shot noise induced fluctuations is reduced by $22 \%$.

Using analytical expressions \ref{eq:maxnoise} and \ref{eq:noise2Q}, the ratio of the noise of the two techniques eventually reads
\begin{equation}% \nonumber
\frac{S_{\theta}}{S_{\theta}^{2Q}} \leqslant 2 \left(\frac{\sigma_{2Q}}{\sigma}\right)^2=\left(\frac{R}{d}\right)^2 \frac{4}{\pi}\left(\frac{1}{C_{max}^{2}}\right)
\end{equation}
Under optimal conditions for both techniques, the numerical value of this ratio is $3.2$, which means that the interferometric technique is a bit noisier than the optical lever technique, when both are perfectly tuned. Nevertheless, our setup offers key advantages over the 2 quadrant detection:
\begin{itemize}
\item \emph{Absolute calibration}: the interferometric measurement only depends on $\lambda$ and $d$, two quantities that can be precisely measured independently, whereas the optical lever sensitivity depends on the exact focalization of the beam and needs to be calibrated for every experiment.
\item \emph{Extended deflection range}: in the present example, deflection up to $10^{3}$ greater can be studied with the quadrature phase interferometric method (and this factor could even be raised by choosing a bigger separation of the analyzing beams in each arms). It implies that strong variations of $\theta$ cannot be studied with great precision with the optical lever detection, for which any slow drift requires a constant adjustment of the $0$. The sensitivity of our technique is moreover constant on the whole range.
\item \emph{Translation insensitive}: the measurement is only sensitive to the rotation we are probing (around $\mathbf{e}_{z}$), and is insensitive to any translation, whereas the other method will sense translation along $\mathbf{e}_{y}$ as well as rotation. Our technique is thus more selective and less sensitive to mechanical vibrations of the setup.
\end{itemize}

\section{Conclusion} \label{section:conclusion}

We have proposed a quadrature phase interferometric technique to measure the angular position of a laser beam. The need of a single passage through a calcite beam displacer to produce the interferences minimizes alignment procedures. The use of polarized beams lets us post process the phase difference to produce 2 output signals in phase quadrature, which finally greatly increases the measurement range at full sensitivity. The accuracy of our experimental realization is $\SI{6.8e-10}{rad}$ ($\SI{1.4e-4}{\arcsec}$) for $\SI{1}{kHz}$ bandwidth on a range of $\SI{20}{mrad}$ ($\ang{1.15}$). This extremely low level is comparable to that of previous studies of fluctuations theorems in our laboratory \cite{Douarche:2005}, but this new setup is easier to handle as it requires a single parallel beam on the rotating mirror instead of two. It could even be improved in various ways (larger calcite and beam size, brighter light source, larger separation of beams in detection area, etc.) Although the accuracy of the optical lever technique may be a bit better than that of the interferometric setup, our technique offers several advantages: robustness (insensitivity to thermal drift, and in general to mechanical vibrations expect for the rotation probed), absolute calibration, large angular range. 

As a final remark, let us point out another way to use the setup: one can rotate the measurement beam displacer BD$_{0}$ with a still laser beam. In this configuration, the sensitivity is unchanged and still described by eq. \ref{eq:sens} (where $\theta$ stand for the angular position of the prism this time), but the range is greatly extended. The limitation is no longer due to the analyzing arms (the focusing lenses will always ensure that the beams fall on their respective photodiodes), but simply to the field of view of the initial calcite. For our setup, a $\SI{0.2}{rad}$ range can be easily be explored. Nevertheless, one should take into account variations of $d$ with $\theta$ in this case, since they are not negligible over such a wide range of measurement.
 
\bigskip

{\bf Acknowledgements}

We thank F. Vittoz and F. Ropars for technical support, and N. Garnier, S. Joubaud, S. Ciliberto and A. Petrosyan for stimulating discussions. This work has been partially supported by contract ANR-05-BLAN-0105-01 of the Agence Nationale de la Recherche in France.

\bibliographystyle{unsrt}
\bibliography{SingleBeam}

\begin{thebibliography}{10}
\bibitem{Electronics}
Current preamplifiers are based on {AD}8067 operational amplifiers, sums and
  substractions on {OPA}606 operational amplifiers, while the divisions are
  based on {AD}734 analog multipier/divider.
\end{thebibliography}

\newpage

\appendix

\section{Optimization of 2 quadrant detection} \label{Appendix}

\begin{figure}
 \begin{center}
	\psfrag{R}[br][br]{$R$}
	\psfrag{w0}[br][br]{$w_{0}$}
	\psfrag{x0}[br][br]{$x_{0}$}
	\psfrag{O}[tl][tl]{$O$}
	\psfrag{Delta}[tl][tl]{$\Delta$}
	\psfrag{w}[br][br]{$w$}
	\psfrag{theta}[cl][cl]{$\theta$}
	\psfrag{2Q}[bc][bc]{2 quadrant}
	\psfrag{PhD}[tc][tc]{photodiode}
	\psfrag{x}[tl][tl]{$x$}
	\psfrag{y}[bl][bl]{$y$}
 \includegraphics{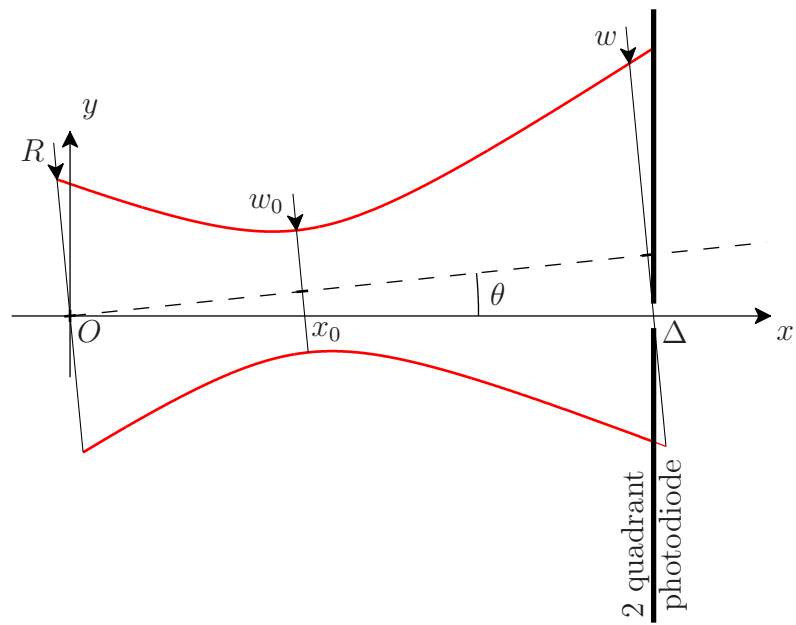}
 \end{center}
 \caption{Principle of the optical lever technique: the incident beam illuminates a 2 quadrant photodiode, and for small deflections $\theta$, the difference between the intensities on the 2 segments is a linear function of $\theta$.}
 \label{fig:principle-2Q}
\end{figure}

In this appendix, we compute the optimal focalization of a laser beam to achieve the best sensitivity in the measurement of a deflexion with a 2 quadrant sensor. The notations are illustrated in Fig. \ref{fig:principle-2Q}: a gaussian beam with $1/e^{2}$ radius $R$ at its origin O makes a angle $\theta$ with the $Ox$ axis. We denote by $x_{0}$ and $w_{0}$ the abscise and radius of the waist of this beam, and $\Delta$ and $w$ the abscise and radius of the beam on the 2 quadrant sensor. $R$ is supposed to be fixed by external constraints (for instance it corresponds to the size of a mirror attached to the rotating object), whereas the focalization of the beam (position of the waist $x_{0}$) and position of the sensor $\Delta$ can be tuned to reach the best sensitivity. The contrast of the intensities of the two segments of the photodiode (ratio of their difference over their sum) can easily be computed as
\begin{equation} \label{eq:contrast-2Q0}
C_{2Q}=\mathrm{erf}\left(\sqrt{2}\frac{\Delta \sin(\theta)}{w}\right)
\end{equation}
where $\mathrm{erf}$ is the error function. 

The best sensitivity $\sigma_{2Q}$ is obtained for $\theta \approx 0$: 
\begin{equation}% \nonumber %\label{eq:contrast-2Q}
\sigma_{2Q}=\left(\frac{dC_{2Q}}{d\theta}\right)_{\theta=0}=\frac{2\sqrt{2}\Delta}{\sqrt{\pi}w}
\end{equation}
Using the properties of gaussian beams, we will now show that the maximum sensitivity is
\begin{equation} \label{eq:sigma-2Q-max}
\max(\sigma_{2Q})=\sqrt{8 \pi}R/\lambda
\end{equation}
and precise how this optimum can be reached. For this purpose, let us define $X$ by
\begin{equation} \label{eq:Xstep0}
X=\sigma_{2Q}\frac{\lambda}{\sqrt{8 \pi}R}=\frac{\lambda \Delta}{\pi R w}
\end{equation}
Eq. \ref{eq:sigma-2Q-max} is thus equivalent to $\max(X)=1$. The radii $R$, $w_{0}$ and $w$ of the gaussian beam for $x=0$, $x_{0}$ and $\Delta$ are linked by the equations :
\begin{eqnarray}% \nonumber %\label{eq:contrast-2Q}
R^{2} & = & w_{0}^2+\left(\frac{\lambda x_{0}}{\pi w_{0}}\right)^{2} \\
w^2 & = & w_{0}^2+\left(\frac{\lambda(\Delta-x_{0})}{\pi w_{0}}\right)^{2}
\end{eqnarray}
We can rewrite these 2 equations as
\begin{eqnarray}% \nonumber %\label{eq:contrast-2Q}
\frac{\lambda x_{0}}{\pi} & = & \pm w_{0} \sqrt{R^{2}-w_{0}^2} \label{eq.:x0step1} \\
\frac{\lambda (\Delta -x_{0})}{\pi} & = & \pm w_{0} \sqrt{w^{2}-w_{0}^2}
\end{eqnarray}
Adding those 2 equations, we immediately get
\begin{equation} \label{eq:Xstep2}
X=\frac{\lambda \Delta}{\pi R w}=  \pm \frac{w_{0}}{R w} \sqrt{R^{2}-w_{0}^2}  \pm \frac{w_{0}}{R w} \sqrt{w^{2}-w_{0}^2}
\end{equation}
Since we're trying to maximize $X$, all $\pm$ signs must be $+$, that is $0 \leqslant x_{0} \leqslant \Delta$. Let us introduce $\alpha$ and $\beta$ in the $[0,\pi/2]$ interval such that $\cos(\alpha)=w_{0}/R$ and $\cos(\beta)=w_{0}/w$, and rewrite equation \ref{eq:Xstep2} as
\begin{eqnarray}
X & = & \frac{w_{0}}{R} \sqrt{1-\frac{w_{0}^2}{w^{2}}}  + \frac{w_{0}}{w} \sqrt{1-\frac{w_{0}^2}{R^{2}}} \\
& = & \cos(\alpha) \sin(\beta) + \cos(\beta) \sin(\alpha) \\
& = & \sin(\alpha+\beta) \label{eq:Xstep3}
\end{eqnarray}
Under this formulation, it is clear that the maximum of $X$ is $1$, so that the maximum of sensitivity of the 2 quadrant measurement is given by Eq. \ref{eq:sigma-2Q-max}. Under optimal conditions, the output of the measurement (Eq. \ref{eq:contrast-2Q0} is thus the following :
\begin{equation} \label{eq:contrast-2Q-Appendix}
C_{2Q}=\mathrm{erf}\left(\sqrt{2}\frac{\pi R \sin(\theta)}{\lambda}\right)
\end{equation}

\bigskip

Let us now precise the conditions of this optimum. According to the definition of $X$ (Eq. \ref{eq:Xstep0}), $X=1$ translate into
\begin{equation} \label{eq:Delta}
\frac{\lambda \Delta}{\pi}=R w 
\end{equation}
The ratio of this last equation with Eq. \ref{eq.:x0step1} leads to
\begin{equation}
\frac{x_{0}}{\Delta} = \frac{w_{0} \sqrt{R^{2}-w_{0}^2}}{R w}
= \frac{w_{0}}{w}\sqrt{1-\frac{w_{0}^2}{R^{2}}} = \cos(\beta)\sin(\alpha)
\end{equation}
According to Eq. \ref{eq:Xstep3}, $X=1$ implies $\alpha+\beta=\pi/2$, hence
\begin{equation} \label{eq:x0step3}
\frac{x_{0}}{\Delta} = \cos(\beta)\sin(\pi/2-\beta)
= \cos^{2}(\beta) = \frac{1}{1+\tan^{2}(\beta)}
\end{equation}
Using Eq. \ref{eq:Delta}, we have
\begin{equation}
\frac{\lambda \Delta}{\pi R^{2}}= \frac{w}{R}=\frac{w}{w_{0}} \frac{w_{0}}{R} = \frac{\cos(\alpha)}{\cos(\beta)} =  \frac{\cos(\pi/2-\beta)}{\cos(\beta)} =  \tan(\beta)
\end{equation}
Introducing $\Delta_{0}=\pi R^{2} / \lambda$, Eq. \ref{eq:x0step3} eventually turns into
\begin{equation}
\frac{x_{0}}{\Delta} =  \frac{1}{1+(\Delta/\Delta_{0})^{2}}
\end{equation}

\begin{figure}
 \begin{center}
 	\psfrag{xl}[Bc][Bc]{$\Delta/\Delta_0$}
 	\psfrag{yl}[Bc][Bc]{$x_0/\Delta$}
 	\psfrag{focused on the sensor}[cl][cl]{\small focused on the sensor}
 	\psfrag{focused at the origin}[cr][cr]{\small focused at the origin}
	\includegraphics{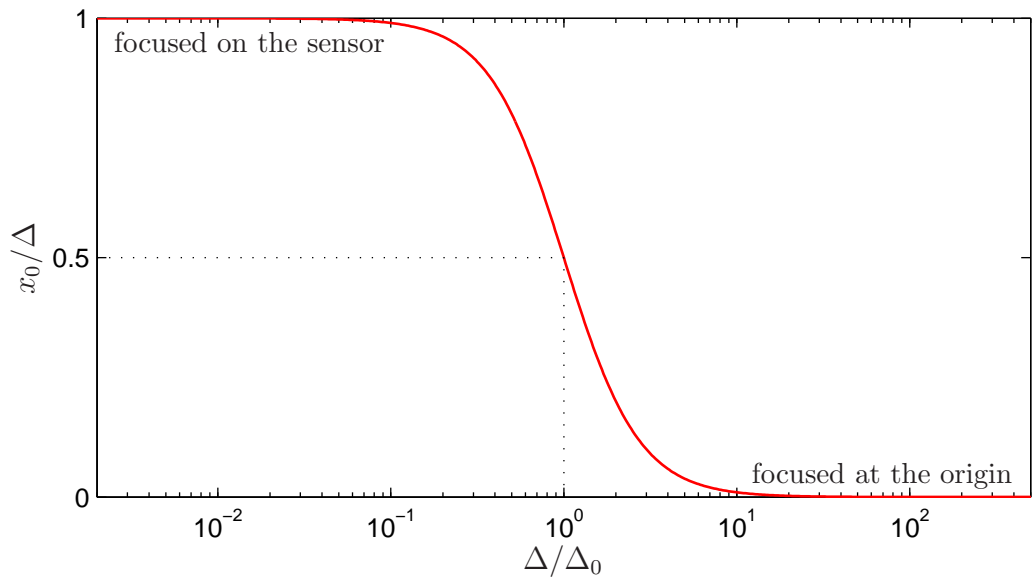}
\end{center}
\caption{Optimal focalization of the beam to achieve best sensitivity in rotation measurement with a 2 quadrant detection. For large values of the distance between the sensor and the axis of rotation ($\Delta \gg \Delta_{0} = \pi R^{2} / \lambda$), one should focus the beam at the origin, whereas for small values of $\Delta$ ($\Delta \ll \Delta_{0}$) it should be focused on the 2 quadrant detector.}
\label{fig:FigAnnexe}
\end{figure}

We plot in Fig.\ref{fig:FigAnnexe} this optimum position of the beam waist as a function of the distance between the origin of the beam and the sensor. There are two limit cases : if $\Delta \ll \Delta_{0}$, the optimum is achieved when the waist is at the origin, whereas if $\Delta \gg \Delta_{0}$ the beam should be focused on the sensor. Let us make a few numerical application for an He-Ne laser to illustrate those limiting cases. If for instance $R=\SI{10}{\micro m}$, we compute $\Delta_{0}\approx\SI{0.5}{mm}$: to probe the deflexion of a small cantilever (typically in an AFM), we most certainly fall in the $\Delta \gg \Delta_{0}$ limit, and the best sensitivity is achieved by focusing the beam on the cantilever \cite{Gustafsson:1994}. On the contrary, if $R=\SI{1}{mm}$, we compute $\Delta_{0}\approx\SI{5}{m}$, and the best practical solution will be to focus the beam on the sensor. Intermediate situations can be found between those 2 limits, with $R\approx\SI{0.1}{mm}$ ($\Delta_{0}\approx\SI{50}{mm}$).

\newpage

\end{document}